\documentclass[acmtoit,acmnow]{acmtrans2m}

\newdef{definition}[theorem]{Definition}
\newdef{remark}[theorem]{Remark}

\usepackage{epsfig}
           
\title{Agent Trade Servers in Financial Exchange Systems}
            
\author{DAVID LYB\"ACK \\ Financial Market Systems, OM AB
	\and MAGNUS BOMAN \\ Swedish Institute of Computer Science (SICS) AB}
            
\begin{abstract} 
New services based on the best-effort paradigm could complement the current deterministic services of an electronic financial exchange. Four crucial aspects of such systems would benefit from a hybrid stance: proper use of processing resources, bandwidth management, fault tolerance, and exception handling. We argue that a more refined view on Quality-of-Service control for exchange systems, in which the principal ambition of upholding a fair and orderly marketplace is left uncompromised, would benefit all interested parties. 
\end{abstract}
            
            
            
            
\begin{document}
            
\begin{bottomstuff} 
Journal Name etc.
\end{bottomstuff}
            
\maketitle

\section{Background}
Exchange operators are now in a business environment shaped by the nature of electronic trading, competing for listings and order flow, and facing the needs of both institutional major traders and retail-level investors. An exchange operation excels when traders display their true willingness to trade. The transaction costs of trading may then shrink, to the benefit of the participants and indirectly also of the marketplace in the competitive environment. 

The electronic exchange is a high-confidence complex system that has to behave in a very well understood and predictable fashion.  This implies that the system must withstand variable load, various failures in supporting systems, faulty user commands, et cetera, and must not by erroneous internal actions cause serious economic damage to the participants of the exchange. While the ideal is continuous matching of simple orders at the equilibrium point with a completely deterministic flow, many needs arise where this ideal cannot be upheld. The complexity of the market entails that decision situations are not repeatable: if a trader makes the wrong choice, very few conclusions can be drawn ex post to improve a similar situation next time, since there is no identifiable next time. The ideal is also under threat from all sorts of failures: hardware, software, peripheral, and network failures. Such failures negatively affect price dissemination routines, market access opportunities, and bid feedback in general. Our starting point is thus not the ideal marketplace, but a real one, the resilience of which is pivotal. 

In a deterministic system, output messages are deterministically repeatable given a stochastic series of input events, i.e. some identified non-deterministic events will completely determine the nature and sequence of the output stream in a well-documented fashion. This implies that if there is any choking in the system whatsoever, the input stream will have to buffer. A best-effort service on the other hand, uses as much resources as it is allocated or until there is no better solution possible. Prolonged execution usually yield ``better'' results for a user, but prolonging is not always an option. The notion of best-effort defines the service result delivered at a particular time as the best one found or best one possible under those given, unrepeatable circumstances, and the service level of the result cannot be impeached: it is rational given the boundary \cite{RuWe91}. 

Introducing the concept of best-effort could open up many new possibilities, but care must be taken so that the maximal downside of using a best-effort service - as opposed to a completely deterministic service - is under control and can be described. It is vital that an add-on service that executes according to best-effort does not in any way compromise the overall ambition of a fair and orderly marketplace. So far, these concerns have hindered best-effort innovations in the financial domain.

We acknowledge the importance of complete tractability in the financial exchange services and therefore propose a hybrid system where new best-effort services complement the current deterministic services. We start in section~\ref{sec:sota} by discussing the state-of-the-art, and thereafter current problems. In section~\ref{sec:solution}, we propose solutions to those problems.  In sections~\ref{sec:advantages} and~\ref{sec:conclusion}, we analyze the advantages and point forwards, respectively.

\section{The State-of-the-Art}
\label{sec:sota}

A strong trend in the software industry is towards Quality-of-Service management with anytime algorithms \cite{DeBo88} for good usage of processing resources and network bandwidth, and clever procedures for graceful degradation whenever there is choking or an outright subsystem failure. An anytime algorithm is a constructive solution to the problem of how to implement best-effort services.

\subsection{Processing Resources}
Contingency orders depend on other orders in the marketplace. Some of the functionality in contingency orders can be provided by the distributed local trading clients, but only the core exchange system can guarantee deals involving more than one trade, such as buying one stock and selling another one simultaneously. Apart from such combinations, one can also place a linked order, i.e., a number of single orders held together by a ``spider object'' with information about exclusive OR-conditions in that set of orders. By using contingency orders, the trading tactic or need is hidden from other participants; only the added liquidity is displayed, to some extent.
A programming technique used for contingency order evaluation is automatic insertion of virtual orders derived from the contingency order. 

\subsection{Bandwidth Management}
Network congestion is (to a reasonable extent) avoided through the use of dedicated leased lines to the exchange system gateways. Custom-built delay boxes are sometimes in place to neutralize systematic latency differences, in order to minimize unfairness between participants. In some systems, the matching service can throttle down automatically (brake the matching service somewhat to reduce dissemination congestion) if the network is choking, to smoothen a temporary peak.

\subsection{Fault Tolerance}
Fault tolerance techniques provide services compliant with the specification after a fault has occured.
For hardware errors, or catastrophic situations like flooding or fire, exchange systems provide hot fail-over mechanisms (based on active replication), i.e. a hot standby replica is ready to take over in a very short time after a serious problem occurred in the primary site. As the secondary has processed the same material as the primary, processing can continue easily (after catch-up and rollback, if needed). Thus, the fail-over procedure is designed to have a minimal impact on users. In fact, users might not even realize that a fail-over has taken place. 

\subsection{Exception Handling}
Exception handling deals with the undefined and unanticipated conditions that, if left unchecked, can propagate through the system and cause a serious fault. Self-verifying software is an ad hoc method used in many  important systems already deployed in our society, including electronic exchange systems, phone switches, and aircraft. A typical exchange system uses a rollback procedure from disk-based transaction logs for a restart, and in problematic cases, the operations staff must identify and (technically) nullify the transaction causing the problem, to enable a successful restart of the system.

Whereas some adaptive rate control procedures are implemented in high-end electronic exchange systems, no system has (to our knowledge) installed components that actively and explicitly base the service offering to the user on a best-effort principle, because one does not want to compromise the paramount paradigm of determinism.

\section{The Problem}
\label{sec:problem}

In this section, we will focus on unsolved problems within the four key areas explained in the previous section. We build on our own experiences of programming real-time trading agents. Parts of this experience comes from participating in the Trading Agent Competition (TAC), see \url{http://tac.eecs.umich.edu/}. Compared to artificial markets, such as the Artificial Stock Market (see \url{http://ArtStkMkt.sourceforge.net/}) or idealized games (see, e.g., \cite{RuMiPa94}), TAC is a realistic, albeit simple, emulation of real agent electronic trading. In order to keep its focus on practical issues, TAC was set up not as a (complex) stock exchange, but as an intelligent assistant competition. The assistants acquire resources for eight different customers on real-time Internet auctions.

\subsection{Processing Resources}
Contingency order matching can easily degrade the performance of the matching service. 
Basic Quality-of-Service control can temporarily suspend the evaluation of certain complex orders to ensure the upkeeping of basic matching service performance at all times. This is a crude maneuver, and will not suffice as users expect their combination orders kept active at all times. Future order types with more conditions and parameters will also increase the expectations on the evaluation policy (cf. \cite{YgAk00}). In effect, users will slowly start to expect a more refined view on Quality-of-Service, like in all areas of Information Technology (cf., e.g., the Open Financial Exchange effort \cite{Co01} in which financial data is exchanged over the Internet in XML). 
The non-deterministic nature of multi-threaded processes with anytime algorithms has so far effectively kept them banned from live exchanges, while they play an important role on artificial marketplaces. For instance, the constraint solver implemented by Mats Carlsson in our TAC'02 agent 006, computes optimal amounts of various goods using an anytime algorithm \cite{Auetal02}.

On the other hand, there will arise temporal series of events in a full-scale, multi-threaded non-deterministic matching service that are practically non-repeatable, since the operating system is involved in the algorithm. For example, a single context switch event may not necessarily occur at exactly the same time or in the same way when replaying a situation, and may thereby promote another flow of events. This inherent property of non-deterministic services makes simulation, verification, and debugging extremely hard. 
This also means that no guarantees of basic performance or temporally dependent completions can be given a priori in a non-deterministic service, which makes it inappropriate for many heavy participants, considering the financial values involved. In other words, one would like best-effort processing services in certain cases, to open up for evaluation of more complex orders, but one can not give up the deterministic stance for the matching engine software.

\subsection{Bandwidth Management}
Congestion that impacts order transactions or crucial price updates will undoubtedly lead to business problems for the participant and eventually for the exchange itself. A systematic latency that is not fair or a partial failure that affects some participants is unacceptable, and may even close the exchange until the problem is resolved. Interactive traffic such as order submission is very sensitive to latency caused by the need to retransmit after a packet loss. Treating packet loss as an indication of congestion in the network as done in TCP is inappropriate for time-critical transactions - such as most order submissions today. 

General throttling is not an ideal solution for bandwidth management in all cases. If more traffic is to become best-effort throughput, there should to be an alternative where there is less business sensitivity to inevitable small delays. One may argue that future networks with Active Queue Management \cite{RaFlBl01} which does not use overflow as a the only indication of congestion, will somewhat reduce the impact of a loss on latency-sensitive flows, but this is far from enough, considering the second-by-second importance of quotes in highly liquid orderbooks. Further, many exchanges prioritize between similar orders based on their arrival timestamp in milliseconds. We therefore see a future need of an explicit rate control, and a refined view of congestion. Furthermore, at the retail level, the drawbacks of TCP/IP are already at hand, since end customers do typically not lease dedicated lines to their respective Internet brokers. 

\subsection{Fault Tolerance}
A fail-over mechanism is often not a sufficient solution, as a logical problem leading to a stop in the primary site will also cause the exact same problem in the secondary site. Redundant replication is the classical way to get reliability from a sub-system, and since it can not be applied to software logic errors, this has created a software problem, which will continue to affect dependability.

\subsection{Exception Handling}
On the one hand, fault-detection checks, which make a whole partition grind to a halt just because of one problematic minor software logic situation somewhere deep in one of many orderbooks, is not the exception handling one would like in a modern software system. On the other hand, no mistakes must be made by the system whatsoever. As mentioned above, it is totally unacceptable that the system would perform erroneous executions in a repeating or even aggravating fashion. The operator prefers a technical stop compared to a long series of faulty deals, for obvious reasons. Therefore the programmer may add invariants at vital points in the code, to defuse any potentially dangerous problem immediately. The obvious problem with self-checking software is its lack of rigor and the unclear coverage. There is always a trade-off between control and performance. Consequently, one method and level of exception handling is not appropriate for all parts of the software. The lack of proper technologies for ultra-reliable systems has created an unsatisfying situation when it comes to exception handling.

In summary, the demand for complete tractability and total determinism puts harsh restrictions on the services and architecture options. We believe that it is the lack of proper assurance techniques that currently keeps ambitions of best-effort execution from the exchange systems arena.

\section{The Solution}
\label{sec:solution}

While the ability to control the rate at which objects are refreshed is a nice tool for ensuring optimal usage of available bandwidth in network-enabled games, it requires good design judgment if best-effort transport via Internet is to be used professionally in active financial marketplaces. We believe that best-effort transport is most reasonable when combined with best-effort execution of scenario-based orders, because of the much lower criticality of its client to front-end server interaction timing. The marketplace operator can consider our suggested technical solutions in a backcasting milieu. 

\subsection{The Agent Trade Server}
To build upon the robustness of the deterministic exchange system platform, we propose a complementary subsystem operating fully under the best-effort principle, both for its bandwidth management, processing resource management, fault tolerance, and exception handling. We acknowledge the importance of continued complete tractability in the current services and therefore propose a hybrid system where the new, optional best-effort services are contained in a separate subsystem: the Agent Trade Server (ATS), complementing the current deterministic services.

A cardinal use case for the new subsystem is when the user uploads a latent order combined with a trigger configuration, combined into a so-called Agent Instruction, to the ATS.
The ATS, operating under the best-effort principle, communicates with the exchange system platform using one of the already available protocols. 

\subsection{Proximity Advantage}
\begin{figure}
\centering
\epsfig{figure=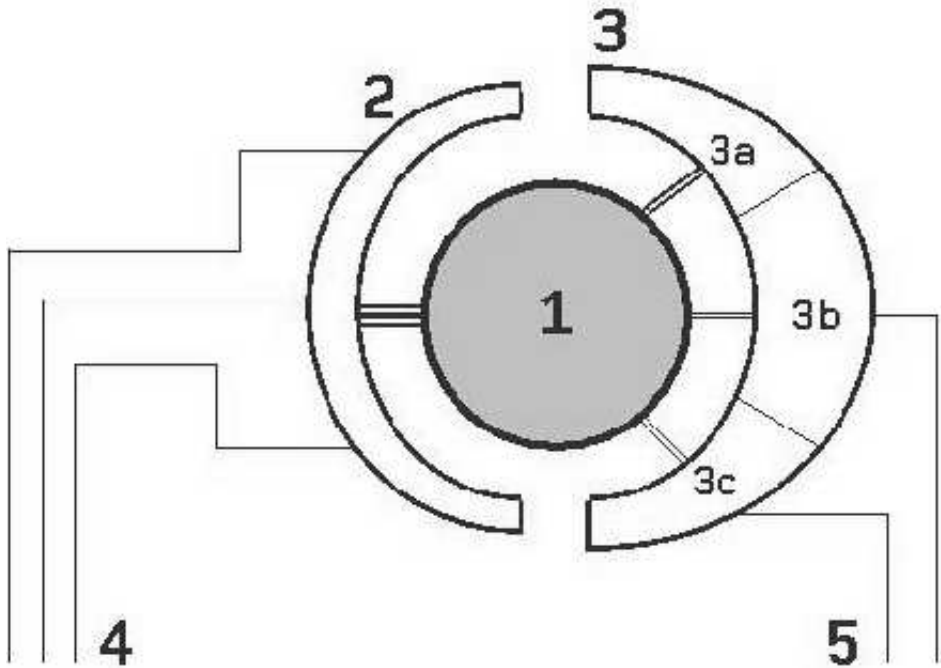, width=12.8cm}
\caption{The Agent Trade Servers setup.}
\label{fig:david}
\end{figure}

The ATS computer will typically be placed physically close to the computer hosting the exchange system back-end processes (core), and between the ATS and the core will be a high-speed local connection (cf. Figure~\ref{fig:david}). The parts comprising the complete setup with Agent Trade Servers are:
\begin{enumerate}
\item Deterministic core with the matching service proper.
\item Multi-protocol gateway(s).
\item Best-effort appurtenance: one or many adjacent servers, running encapsulated complementary best-effort services. The appurtenance is split into three parts:
	\begin{enumerate}
	\item Core services extension(s), e.g., a computer running Combinatorial Solution Finder, or other supportive services.
	\item Agent Trade Server(s) type I, with user-configurable agents spawned under control of a dispatch service, utilizing multi-threading or similar non-verified resource sharing under the operating system of the server.
	\item Agent Trade Server(s) type II, with user-programmable agents executing under sandbox resource management, or similar sharing mechanisms. There is extensive shielding between individual agents, so that one erroneously programmed or extra demanding agent cannot affect others in an uncontrollable way.
	\end{enumerate}
\item External connections via gateway(s), typically through either leased lines or shared IP links. Using these connections, users run interactive sessions for order submission, and subscribe to multi-cast messages, such as price updates.
\item External connections to the best-effort appurtenance, typically via the Internet for the interactive sessions and the subscriptions of, for example, status updates. Note that servers in the appurtenance classified as 3a above do not have external connections, but only internal links with the core.
\end{enumerate}

\subsection{Agent Trade Server Operation}
The ATS will, upon receiving an agent instruction, create a new thread to handle the request, or revise a previously spawned thread or process. A precondition for the automatic activation of a latent order in the ATS is a Boolean operation on a set of scenario triggers. Inspirational generic scenarios can be proposed by the marketplace operator, and then adjusted by each trader or investor. The scenarios could be saved in personalized galleries, and optionally calibrated for each individual order. The agent in the ATS will continuously evaluate the triggers in its instruction against real-time market information received from the core.

To avoid resonance effects, the bourse operator would typically not allow only one short-term fluctuation in an index value as a single trigger. Instead, the trader should set a scenario by combining at least two (not totally correlated) triggers. Note that a real-time scenario trigger must evaluate the current market situation within specified timing constraints, quite short-termed, possibly slightly trading-off the quality of its own analysis if needed. Thus, even the trigger in itself is under Quality-of-Service control.

The user could be allowed to set a minimal acceptable quality of the best-effort service for each individual agent instruction (according to a measurement yet to be defined), below which the agent is not allowed to trade. Another challenge is to investigate what scenarios that a trader would like to trigger some latent orders, and how the quality of real-time triggering is to be measured.

The operator would also be constrained with respect to integrity and security issues. The trust relationship between marketplace and customer will be put to harsh tests, as the customer in the typical case is reluctant to reveal her preferences. As most uploads encode true preferences, this information is ideally processed without human intervention, and must be treated as maximally sensitive data. 

The drawback for the trader with communicating trade intention early, can be more than offset by the inherent premium in the reliable, cheap, high-speed connection between the core and the ATS. The proximity and direct connection is effectively feeding the ATS with market data much faster than anyone else, thus putting the Agents in the ATS at an advantage in certain cases. This may be a strong enough incentive to stimulate an early submission of orders via the ATS - when appropriate.

\subsection{Agent Programming}
Although we primarily envision a few archetypical agents to be offered by the exchange for traders to configure, it is not unreasonable in later generations of the system to open up so-called ``sandbox areas'' for user-created code to execute in. To encode trade logic into agents means that the code can be pro-active, autonomous, and encapsulated. These features make agents an interesting option for future complex order types. The user can also make it hard for unauthorized persons to grasp the preferences encoded in the agent, and make it virtually impossible to apply reverse engineering techniques. The programming tools for creating agents are increasing in number and quality, as is the number of agent programming consultants and brokers, and plenty of code is shared through online communities. Academic research has pursued this track for years, also as connected to a marketplace scenario. Examples of our own recent efforts in this field include agent-based simulations on real stock market data \cite{BoJoLy01}, and multi-agent systems situations, in which several artificial agents operate simultaneously on the same marketplace, subject to norms and other constraints \cite{BoBrHaJaKuVe01}. 
Although there is substantial previous work in the academic field when it comes to agent-mediated commerce, we feel that it is often unrealistic in the way it commonly approaches agent-mediated trading \cite{WeGo99} and agent mobility \cite{Ru99}. The agency gateway is a multiprotocol connectivity zone to reach users via various end-user devices. At SICS, ways of providing accessible APIs for end-user services have been produced \cite{By01}. International community-shaping efforts, such as .NET and the possibility of SOAP adaptation for web services are also likely to be important in the near future. 

It should be noted that agent/server communication also has its own set of problems. An agent server is under constant threat of overload, due to malicious intent or bad agent code. In the TAC server, agent requests which are looped without correct termination conditions produce problems. It is also difficult for the people responsible for the server to shut out agents due to erroneous behavior, since the reasons for the shutout should be communicated to the agent owners, which is sometimes impossible. As a result, agent owners might find their agent has not been running as expected, and this without notification, let alone explanation. In a financial exchange, this is totally unacceptable. In addition, a study made in the first phases of programming our TAC'01 agent 006 \cite{HaOr02} analyzes the following latency concerns.
\begin{itemize}
\item Round trip time for agent- or server-generated message
\item Receive interval (from message receipt start until finish)
\item Intrinsic data time bound (e.g., a bound on the validity of message data)
\item Agent deliberation time (e.g., for interpreting a message)
\item Throughput: number of messages per time unit or clock cycle
\item Time-slotting: agent planning activities for sending and receiving messages
\item Network topology, affecting message delivery time
\end{itemize}

These concerns should be added to the usual performance considerations \cite{Ta96}, including:
\begin{itemize}
\item Traffic congestion caused by temporary resource overload
\item Resource imbalance (e.g., between network and processor capacity)
\item Broken cables 
\item Broken connections (e.g., overloaded hosts or routers)
\end{itemize}

Moreover, the sheer amount and varibility of the code running on an agent server makes code verification a practical impossibility. To get a crude feel for the amount of code expected, consider that our TAC'01 agent RiskPro consisted of about 7000 lines of code, 3000 lines of which were devoted to communication with the TAC server, over TCP/IP \cite{Bo01}. There is currently considerable effort being put into solving these problems, however, both in the form of market-driven and in the form of purely academic research.

\section{The Advantages}
\label{sec:advantages}

\subsection{Processing Resources}
New scenario-based orders can be evaluated continuously in the ATS without degrading the basic matching service. This means that the exchange system as a whole can offer not only the matching itself, but also resources for tactical trading without negative impact on the basic services. Special limitations surround combination orders. Only the core can fully match a combination, but cumbersome combinations should be kept away from the core as long as possible. Moreover, instead of a large combination, the ATS can formulate a basket of smaller combinations in a quick sequence, accepting small residuals. That said, the major advantage of an ATS when it comes to processing resources, is the continuous scenario evaluations ongoing in the agents. Only when a latent order triggers, it will be formulated in detail and submitted to the core. The user could be allowed to set a minimal acceptable quality of the best-effort matching service for each individual order, according to a measurement yet to be defined, below which the service is automatically disregarding that order. Another challenge is to investigate how the quality of real-time triggering is to be measured. 

\subsection{Bandwidth Management}
We argue that regardless of how clever future transport protols may become, best-effort transport solutions between the marketplace and the end user are most reasonable when combined with best-effort evaluation of scenario-based orders, such as in the ATS, because of the much lower criticality of its client/server interaction timing. 

Furthermore, the more intentions that could be communicated in advance of unfolding market events, the more the network load smoothens, which reduces the need for expensive reserve capacity in the network. Also, traffic between the ATS and the core would not be hindered, or slowed down, by network load in the Wide Area Network or the Internet, due to the direct or LAN connection used. A setup with an ATS may therefore reduce gateway and network loads slightly, which is especially beneficial when approaching heavy load situations. Throttling may then in certain situations be eluded. This in turn might slightly increase the number of completed deals per minute under those circumstances.

\subsection{Fault Tolerance}
Keeping as much as possible of the new functionality out of the deterministic core altogether will drastically reduce the risk of introducing new code into the heart of the system. Containment of new functionality away from the core is in itself a strategy for upkeeping and increasing highest possible reliability while the system is under continuous development. A crashed ATS thread was presumably only evaluating a single agent instruction from one user, and can therefore be out of action for a few seconds (or even die) without widespread consequences, whereas the core in a sense is the marketplace, and must not fault. 
The best-effort nature of the ATS allows for many possibilities in a fail-over situation to degrade its services until fully recovered to normal operating status. A recovery unit can replace a failed process, and concentrate on agent restart based on the user-submitted agent instruction, and can discard processing states. 

\subsection{Exception Handling}
Error containment and exception handling will be in effect locally in the ATS, and internal problems are not to affect the core services.

\section{Conclusion}
\label{sec:conclusion}

We have suggested an Agent Trade Server as a complement to an exchange system, without compromising the core service, and so:
\begin{itemize}
\item open up for proper best-effort transport connectivity, fully in line with Internet design foundations and contemporary trends in the industry
\item prepare for the introduction of new order details and tactical trading instructions
\item provide a new level of containment to better defend the complex system from fatal software error propagation when new services are introduced in the future.
\end{itemize}

\section*{Acknowledgements}
The authors would like to thank Johan Kummeneje and Lars Rasmusson for draft comments. Lyb\"ack was supported through his research assignment at OM, while the VINNOVA project TAP on accessible autonomous software provided Boman with the time required for this study.


\bibliographystyle{acmtrans}
\bibliography{toitm4}
\begin{received}
Received March 2002;
revised...; accepted...
\end{received}
\end{document}